\documentclass[conference]{IEEEtran}
\usepackage{array}
\newcolumntype{P}[1]{>{\centering\arraybackslash}p{#1}}

\usepackage{forest}

\usepackage{listings}
\usepackage{xcolor} %
\usepackage{multirow}
\usepackage{booktabs} %
\usepackage{amsmath}
\usepackage{amssymb}
\usepackage{graphicx}

\usepackage{subcaption}
\usepackage{rotating}
\usepackage{tabularx}
\usepackage{wrapfig}
\usepackage{listings}
\usepackage{color, colortbl}
\usepackage{balance} %
\usepackage{lscape}
\usepackage[ruled,linesnumbered]{algorithm2e}
\SetKw{KwStep}{Step}
\usepackage{hyperref}
\usepackage{xspace}
\usepackage{framed}
\usepackage{syntax}
\usepackage{comment}
\usepackage[flushleft]{threeparttable}
\usepackage{forest}
\usepackage{pifont}
\usepackage{amsfonts}
\newcolumntype{C}[1]{>{\centering\arraybackslash}p{#1}}

\usepackage[framemethod=TikZ]{mdframed}
\usepackage{soul}
\usepackage{xcolor}

\usepackage[most]{tcolorbox}
\usepackage{flushend}
\usepackage{enumitem}
\usepackage[normalem]{ulem}
   
\usepackage{tikz}

\usepackage[numbers,sort&compress]{natbib}

\newcommand*\rot{\rotatebox{90}}

\definecolor{codegreen}{rgb}{0,0.6,0}
\definecolor{codegray}{rgb}{0.5,0.5,0.5}
\definecolor{codepurple}{rgb}{0.58,0,0.82}
\definecolor{backcolour}{rgb}{0.95,0.95,0.92}
\definecolor{Gray}{gray}{0.1}
\lstdefinestyle{mystyle}{
	backgroundcolor=\color{backcolour},   
	commentstyle=\color{codegreen},
	keywordstyle=\color{magenta},
	numberstyle=\tiny\color{codegray},
	stringstyle=\color{codepurple},
	basicstyle=\scriptsize,
	breakatwhitespace=false,         
	breaklines=true,                 
	captionpos=b,                    
	keepspaces=true,                 
	numbers=left,                    
	numbersep=5pt,                  
	showspaces=false,                
	showstringspaces=false,
	showtabs=false,                  
	tabsize=2
}

\lstdefinelanguage{Pythonna}{%
	language     = python,
	morekeywords = {to_categorical, flow_from_directory, pad_sequences, load_image}
}

\lstdefinestyle{customc}{
	belowcaptionskip=1\baselineskip,
	breaklines=false,
	frame= single,
	breaklines = true,
	xleftmargin=\parindent,
	language= Pythonna,
	showstringspaces=false,
	basicstyle=\footnotesize\ttfamily,
	keywordstyle=\bfseries\color{green!40!black},
	commentstyle=\itshape\color{purple!40!black},
	identifierstyle=\color{blue},
	stringstyle=\color{codegreen},
	backgroundcolor=\color{gray!4}
}

\graphicspath{{./figures/}}

\newcommand\tool{\textsc{$\mu$PRL}\xspace} 

\newcounter{rqs}
\stepcounter{rqs}

\newcounter{NumObservations}
\stepcounter{NumObservations}
\definecolor{shadecolor}{rgb}{.9,.9,.9}

\newcommand{\head}[1]{\noindent\textbf{#1.}}

\newenvironment{rltaxonomy}
{
\forest
  for tree={
    font=\sffamily\small, %
    grow'=0,
    child anchor=west,
    parent anchor=south,
    align=left, base=bottom,
    anchor=west,
    calign=first,
    inner sep=0,
    edge path={
      \noexpand\path [draw, \forestoption{edge}]
      (!u.south west) +(7.5pt,0) |- node[fill,inner sep=1.25pt] {} (.child anchor)\forestoption{edge label};
    },
    before typesetting nodes={
      if n=1
        {insert before={[,phantom,  align=left, base=bottom]}}
        {}
    },
    fit=band,
    before computing xy={l=15pt},
   }
}
{
\endforest
}

\AtBeginDocument{%
  \providecommand\BibTeX{{%
    \normalfont B\kern-0.5em{\scshape i\kern-0.25em b}\kern-0.8em\TeX}}}

\begin{document}

\title {\tool: A Mutation Testing Pipeline for Deep Reinforcement Learning based on Real Faults}

\author{\IEEEauthorblockN{
		Deepak-George Thomas \IEEEauthorrefmark{1},
		Matteo Biagiola \IEEEauthorrefmark{2}, 
		Nargiz Humbatova \IEEEauthorrefmark{2},
		Mohammad Wardat \IEEEauthorrefmark{3},
        Gunel Jahangirova \IEEEauthorrefmark{4}, \\
        Hridesh Rajan \IEEEauthorrefmark{5},
        Paolo Tonella \IEEEauthorrefmark{2}
	} \vspace{0.5em}
	\IEEEauthorblockA{\IEEEauthorrefmark{1} Dept. of Computer Science, Iowa State University, Iowa, USA, dgthomas@iastate.edu}
	\IEEEauthorblockA{\IEEEauthorrefmark{2} Universit\`a della Svizzera italiana, Lugano, Switzerland, 
		\{matteo.biagiola, nargiz.humbatova, paolo.tonella\}@usi.ch}
	\IEEEauthorblockA{\IEEEauthorrefmark{3} Dept. of Computer Science and Engineering,  Oakland University, Michigan, USA, wardat@oakland.edu}
 	\IEEEauthorblockA{\IEEEauthorrefmark{4} King's College London, London, United Kingdom, gunel.jahangirova@kcl.ac.uk}
  	\IEEEauthorblockA{\IEEEauthorrefmark{5} School of Science and Engineering,  Tulane University, Louisiana, USA, hrajan@tulane.edu}
  
}

\maketitle
\thispagestyle{plain}
\pagestyle{plain}

\begin{abstract}

Reinforcement Learning (RL) is increasingly adopted to train agents that can deal with complex sequential tasks, such as driving an autonomous vehicle or controlling a humanoid robot. Correspondingly, novel approaches are needed to ensure that RL agents have been tested adequately before going to production. Among them, mutation testing is quite promising, especially under the assumption that the injected faults (mutations) mimic the real ones.

In this paper, we first describe a taxonomy of real RL faults obtained by repository mining. Then, we present the mutation operators derived from such real faults and implemented in the tool \tool. Finally, we discuss the experimental results, showing that \tool is effective at discriminating strong from weak test generators, hence providing useful feedback to developers about the adequacy of the generated test scenarios.

\end{abstract}

\begin{IEEEkeywords}
    reinforcement learning, mutation testing, real faults
\end{IEEEkeywords}

\section{Introduction}
\label{sec:intro}

Reinforcement Learning (RL) is being applied to various safety-critical systems such as traffic control, drone navigation, and power grids~\cite{liu2023traffic,chauhan2023powrl, kaufmann2023champion}. Due to its relevance in such systems, RL developers need to make sure that their RL agent is tested thoroughly. Mutation testing~\cite{demillo1978hints,Hamlet1977TestingPW} is a powerful technique to ensure that the test set exercises the system in an adequate way, but existing attempts to apply mutation testing to RL~\cite{LuRLMut, TambonRLMut} are limited, and do not cover the full spectrum of faults that may affect an RL agent. 
In this paper, we construct a comprehensive taxonomy of real faults identified by repository mining (we analysed 2,787 posts, 3.6 times more  than previous work~\cite{nikanjam2022faults}) and we develop an RL mutation tool, named \tool, which implements new mutation operators (MOs) mimicking the real faults in the taxonomy.

Our taxonomy of real RL faults targets RL developers who use well-known frameworks, such as StableBaselines3~\cite{stable-baselines3} and OpenAI Gym~\cite{1606.01540}, for their projects. %
We mined StackExchange and GitHub posts and then manually analysed the relevant artifacts to identify real bugs that developers experience.
Then, we derived 15 mutation operators from the taxonomy and implemented 8 of them in the tool \tool. %
These operators have been evaluated on four environments provided by OpenAI Gym~\cite{1606.01540} and HighwayEnv~\cite{LeurentAnEnvironment2018}: CartPole, Parking, LunarLander, and Humanoid~\cite{barto1983neuronlike, tassa2012synthesis, lunarlander, LeurentAnEnvironment2018}. The Humanoid environment is a particularly challenging robotics environment with a high dimensional observation space, thereby requiring extensive computational resources. In the evaluation, we applied \tool to environments with both discrete and continuous action spaces, and we considered popular Deep RL (DRL) algorithms, including DQN, PPO, SAC, and TQC~\cite{mnih2015human, schulman2017proximal, haarnoja2018soft, kuznetsov2020controlling}. %

Experimental results indicate that our mutation tool \tool is useful in discriminating strong from weak test scenario generators and achieves a high sensitivity  to the test set quality, substantially higher than that of the state of the art tool RLMutation \cite{TambonRLMut}. We also show that the new fault types introduced in our taxonomy are major contributors to the increased sensitivity of our mutation operators.

\section[Related]{Related Work} \label{sec:relatedwork}

We organise the related works into those analysing Deep Learning (DL)/RL faults and those mutating DL/RL models.

\subsection{Fault Classification}
\textbf{DL Faults:} 
Humbatova \etal~\cite{humbatova2020taxonomy} proposed a DL faults taxonomy using StackOverflow and GitHub artifacts. %
Islam \etal~\cite{islam2019comprehensive} studied the frequency, root cause, impact and pattern of bugs while using deep learning software. %

\textbf{RL Faults:} Nikanjam \etal~\cite{nikanjam2022faults}  developed a fault taxonomy for RL programs. They studied 761 RL artifacts obtained from StackOverflow and GitHub. While analysing GitHub they went over issues from the following frameworks - OpenAI Gym, Dopamine, Keras-rl and Tensorforce. 

Their work focuses on mistakes developers make while writing an RL algorithm from scratch. However, RL algorithms are notoriously hard to implement~\cite{henderson2018deep}, and even small implementation details have a dramatic effect on performance~\cite{engstrom2020implementation,ppo-37-details}. Our work considers the perspective of software developers who use RL as a tool to address an engineering problem. Therefore, we focus on bugs that arise while using reliable RL implementations~\cite{castro18dopamine, 1606.01540, tensorforce, plappert2016kerasrl, nikanjam2022faults, stable-baselines3} 
(or bugs that can be mapped to those occurring in reliable implementations)
. We compare with the taxonomy of Nikanjam \textit{et al.}~\cite{nikanjam2022faults} in \autoref{sec:cmp}.

Andrychowicz \etal~\cite{andrychowicz2021matters} studied the effect of more than 50 design choices on an RL agent's performance by training around 250$k$ agents. 
They used this study to recommend various hyperparameter choices. %

\subsection{Mutation Testing for AI-based systems} \label{sec:mutation-testing}

\textbf{Mutation testing for DL:}
DeepMutation~\cite{deepmutation} and MuNN~\cite{munn}, were the pioneers in recognising the need for mutation operators specifically tailored to DL systems. Subsequently, DeepMutation was extended to a tool called DeepMutation++~\cite{hu2019deepmutationpp}, focusing on operators that can be applied to the weights or structure of an already trained model. Jahangirova and Tonella~\cite{JahangirovaICST20} performed an extensive empirical evaluation of the mutation operators proposed in DeepMutation++ and MuNN. 
DeepCrime~\cite{deepcrime} differs from DeepMutation++ in that it uses a set of mutation operators derived from \textit{real faults} in DL systems~\cite{humbatova2020taxonomy}. Such operators are applied to a DL model before training.

\textbf{Mutation testing for RL:}
There are currently two papers dedicated to mutation testing of RL systems. Lu \etal~\cite{LuRLMut} introduce a set of RL-specific mutation operators, including \textit{element-level} mutation
operators and \textit{agent-level} mutation operators. The element-level mutations consider the injection of perturbations into the states and rewards of an agent. 
Agent-level mutations introduce errors into an already trained agent. If a trained agent's policy is represented by a Q-table~\cite{sutton2018reinforcement}, an agent-level mutation would fuzz the state-action value estimations stored in the table. 
If the policy of a trained agent is implemented by a neural network, an agent-level mutation would remove a neuron from the input or output layer of the network.
In addition, the authors propose operators that affect the exploration-exploitation balance by, for instance, mutating the exploration rate of the agent during training.

However, such mutation operators are not based on any real-world fault taxonomy or existing literature. Moreover, mutation killing is computed on a single repetition of the experiment, not accounting for randomness in the training process. Previous literature shows that RL algorithms are sensitive to the random seed~\cite{henderson2018deep}, suggesting that statistical evaluations are needed to draw reliable conclusions~\cite{rishabh2021edge}.

Tambon \etal~\cite{TambonRLMut} explore the use of higher order mutants, i.e., the subsequent application of different mutations to a program under test, in the context of RL. The mutation operators they propose, implemented in the tool RLMutation, are based on a number of sources. As a basis, the authors have adopted the existing operators for RL~\cite{LuRLMut} and DL systems~\cite{hu2019deepmutationpp, munn, deepcrime} and complemented them with operators extracted from the state-of-the-art taxonomies of RL faults~\cite{nikijamTaxonomy} and real DL faults~\cite{humbatova2020taxonomy}. They divided the obtained operators into three broad categories: \textit{environment-level}, \textit{agent-level}, and \textit{policy-level} operators. Mutations at the \textit{environment-level} include faults related to the observations the agent perceives in the environment, for instance due to faulty sensors or deliberate attacks.  
Operators at the \textit{agent-level} stem from the faults that developers make while implementing RL agorithms, such as omitting the terminal state or selecting a wrong loss function. Finally, \textit{policy-level} mutations focus on the agent's policy network, mutating activation functions or  number of layers. These mutations are used to create first-order mutants. Among them, those that are not trivial, i.e., that are not killed by all the test environments, are considered for  higher-order mutation. 

Our mutation tool \tool differs substantially from both Lu \etal~\cite{LuRLMut} and Tambon \etal~\cite{TambonRLMut}, because it is grounded on a novel taxonomy of real faults experienced by developers that rely on existing, mature frameworks for the creation of RL agents. This rules out the syntactic state/reward/policy perturbations introduced by Lu \etal~\cite{LuRLMut} and the mistakes made by programmers when implementing RL algorithms from scratch that are instead considered by Tambon \etal~\cite{TambonRLMut}. Since RLMutation~\cite{TambonRLMut} includes also real faults that may occur when developers rely on existing RL frameworks, we conduct a detailed comparison with the existing taxonomy and tool in \autoref{sec:cmp} and \autoref{sec:mut-tool-eval-res},  respectively.

\section{Taxonomy of real RL faults} \label{sec:Approach}

We constructed a taxonomy of real RL faults in a bottom-up way, starting from the collection of artefacts, obtained through software repository mining. We then labeled such artefacts, to eventually organise the labels into a taxonomy.

\subsection{Mining of Software Artefacts}

In our preliminary investigation, we observed that most discussions about faults reported by developers implementing an RL agent happen in Stack Exchange. We also noticed several commit messages about RL faults in GitHub. Hence, we mined these two repositories.

\subsubsection{Mining GitHub}
The foremost challenge while mining GitHub repositories was identifying popular RL frameworks. Nikanjam \etal used Keras-rl, Dopamine, Tensorforce, and OpenAI Gym \cite{nikanjam2022faults}. However, OpenAI Gym is only used to simulate the interactions between an agent and the environment, and does not provide RL algorithms. While investigating the remaining frameworks we found that Tensorforce is no longer  maintained and Keras-rl did not get updated since November 2019~\cite{castro18dopamine, 1606.01540, tensorforce, plappert2016kerasrl}. Therefore, to identify popular RL frameworks we checked top-tier Machine Learning and Software Engineering Conferences, such as ICML, ICSE, ESEC/FSE, and ASE. We manually inspected 89 papers, dropping all papers that focused on model-based RL, inverse RL, and multi-agent RL. This filtering step left us with 9 papers from SE conferences and 47 papers from ICML-22 that provided actionable insights. While many were custom implementations of RL algorithms, the majority of the papers that used frameworks used StableBaselines3~\cite{stable-baselines3} (7 overall).

We followed Humbatova \etal's~\cite{humbatova2020taxonomy} approach to mine GitHub repositories. We searched for files containing the string ``stable\_baselines3'' using the GitHub Search API, and found 27,413 files. Since the API has a limit of 1$k$ results per query, we searched for file sizes between 0 and 500$k$ bytes, with a step size of 250~\cite{humbatova2020taxonomy}. 
We identified 4,272 repositories corresponding to these files. We then dropped all repositories that had less than 10 stars, 100 commits, 10 forks, and 5 contributors, which left us with 67 repositories. 
As the next step, we manually verified whether they were related to RL and dropped the repositories containing tutorials and code examples using StableBaselines3, obtaining the list of the final 43 repositories. These 43 repositories were then used to extract issues, pull requests (PRs) and commit messages. While extracting the issues, we only extracted those that contained the label ``bug'', ``defect'' or ``error'' in them. Following Islam~\etal~\cite{islam2019comprehensive} we only selected commits that contain the term ``fix''.

To automate the process of extracting relevant artifacts from GitHub, we followed Humbatova \etal~\cite{humbatova2020taxonomy}: we combined all issues, PR titles, descriptions, and comments along with commit messages into a text dump. We did data-cleaning on the words within the text dump (dropped stop words, non-word characters, etc) and counted the frequency of each remaining word. Words that had a frequency lower than 20 (raised from 10~\cite{humbatova2020taxonomy}, to obtain a manageable list of words) were dropped, resulting in a list of 14,921 words. We divided the final list of words among 3 authors to identify relevant RL words and obtained 118. We then selected the corresponding issues, PRs, and commits, a total of  1,120~\cite{islam2019comprehensive, humbatova2020taxonomy}.

\subsubsection{Mining Stack Exchange}

To include  questions on Data Science and Artificial Intelligence, which might be relevant for our taxonomy, we mined both StackOverflow (SO) and Stack Exchange's (SE) Data Science (DS) and Artificial Intelligence (AI) Q\&A websites (with SO falling under the umbrella of SE).

\begin{table}[ht]
	\centering
    \caption{Number of unique tags and posts in SE and SO}
    \label{table:taxonomy:stats}
    \setlength{\tabcolsep}{2pt}
    \renewcommand{\arraystretch}{1}

        \begin{tabular}{r@{\hskip 1em}ccc}
            \toprule

            \multicolumn{1}{l}{} & 
            \textbf{\# Tags (All)} & \textbf{\# Tags (RL)} & \textbf{\# Posts} \\

            \midrule
            
            Artificial Intelligence (AI-SE) & 974 & 104 & 783 \\
            Data Science (DS-SE) & 668 & 9 & 245 \\
            StackOverflow (SO-SE) & 63,653 & 19 & 3,682 \\

            \midrule
            
            \textit{Total} & 65,295 & 132 & 4,710 \\

            \bottomrule
            
        \end{tabular}
\end{table}

We used SE's Data Explorer to extract posts from SO and SE. %
We first extracted all tags from these websites; then, we filtered all tags without the term ``reinforcement'' in their respective name, excerpt (i.e., short description), or wikibody (i.e., detailed description). The resulting 132 tags were then used to select all posts from SE. Next, we excluded all posts without an accepted answer~\cite{humbatova2020taxonomy}. We also filtered out the posts whose title contained the terms ``how'', ``install'' or ``build'', to discard how-to questions. During manual inspection, we found that many posts were not RL specific, but rather related to Machine Learning in general. Therefore, we dropped all posts that had only one tag and it was ``machine-learning''. Following this procedure, we obtained 4,710 posts (see \autoref{table:taxonomy:stats}).

Given the large number of posts, we performed various pilot studies to gauge the data quality of selected samples, and manually filtered out irrelevant posts. These pilot studies yielded a large number of false positives. Upon a closer examination of the dataset, we found that the posts from SO contained around 78\% of the total posts and the tag ``artifical-intelligence'' was present in 2,779 SO posts (without dropping duplicates). A big chunk of the posts within this category contained posts concerning classical AI, such as the A$^*$ algorithm. Therefore, we dropped all posts that either contained the tag ``artificial-intelligence'' or a combination of ``artificial-intelligence'' and ``machine-learning'', without another RL tag. Lastly, following Islam \etal ~\cite{islam2019comprehensive}, we kept only posts that contained code. This brought our final dataset size down to 1,667. %

\subsubsection{Manual Labelling}
One group of labellers manually analysed the artifacts and dropped false positives. Five authors participated in the labeling process for taxonomy construction. Each post in the dataset was randomly assigned to two authors. We used the procedure by Humbatova \etal~\cite{humbatova2020taxonomy} for the labeling process. The authors therein, developed a tool that helped them manage the labels. This tool allowed each assessor to pick a label generated by their colleagues. In case none of the labels matched the post description, they created their own label and added it to the tool, which then became accessible to others. We pre-loaded all labels created by Nikanjam \etal~\cite{nikanjam2022faults} into the labeling tool, to be consistent with, and build upon, the existing RL fault taxonomy~\cite{nikanjam2022faults}. Furthermore, to check the disagreement between various participants, we measured Krippendorf’s Alpha~\cite{krippendorff1970estimating, krippendorff2018content}, which handles more than two raters, with each rater only labeling a subset of the posts. 
Krippendorf proposed to reject data where the confidence interval of the reliability falls below 0.667. Ideally the value of alpha should be 1.00 but variables with values greater than 0.800 could be relied upon~\cite{krippendorff2018content, hayes2007answering}. During labeling, the two raters of each post met together to resolve conflicts, when any such conflict arouse. When no resolution between two raters could be reached for a certain post, the overall group made the final decision through voting~\cite{humbatova2020taxonomy}. 
While the average agreement (Krippendorf’s Alpha) was 0.546 before the consensus meeting, it raised to 0.926 after the meeting.

Finally, all the authors went through all the posts together for a final pass. 

\begin{figure*}
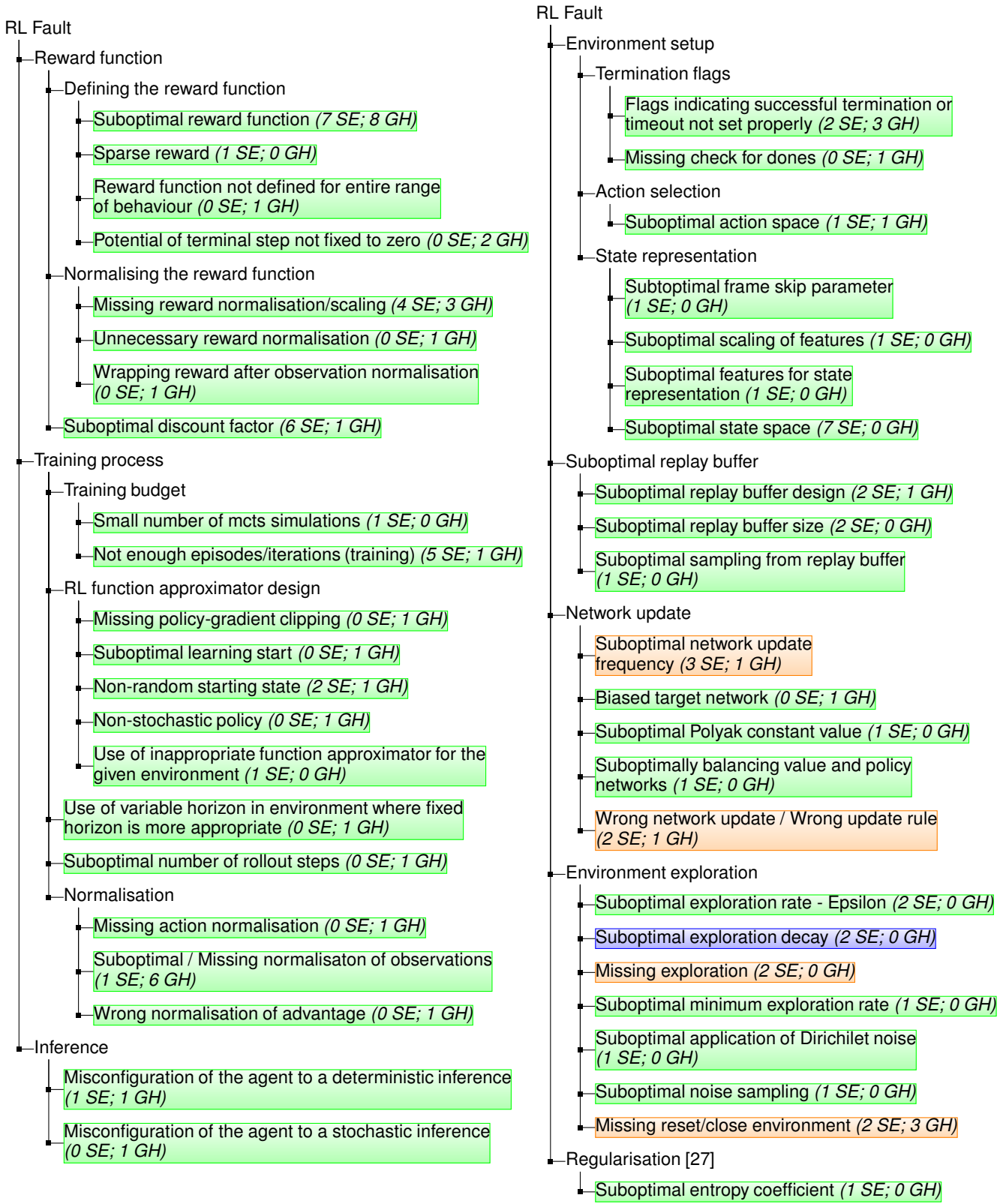

\begin{tabular}{ll}
\begin{minipage}{9cm}
\begin{rltaxonomy}
[RL Fault 
    [Reward function 
        [Defining the reward function
            [Suboptimal reward function \textit{(7 SE; 8 GH)}, draw=green, top color=green!5, bottom color=green!30]
            [Sparse reward \textit{(1 SE; 0 GH)}, draw=green, top color=green!5, bottom color=green!30]
            [Reward function not defined for entire range \\ of behaviour \textit{(0 SE; 1 GH)}, draw=green, top color=green!5, bottom color=green!30]
            [Potential of terminal step not fixed to zero \textit{(0 SE; 2 GH)}, draw=green, top color=green!5, bottom color=green!30]]
        [Normalising the reward function
            [Missing reward normalisation/scaling \textit{(4 SE; 3 GH)}, draw=green, top color=green!5, bottom color=green!30]
            [Unnecessary reward normalisation \textit{(0 SE; 1 GH)}, draw=green, top color=green!5, bottom color=green!30]
            [Wrapping reward after observation normalisation \\ \textit{(0 SE; 1 GH)}, draw=green, top color=green!5, bottom color=green!30]]
        [Suboptimal discount factor \textit{(6 SE; 1 GH)}, draw=green, top color=green!5, bottom color=green!30]
        ]
    [Training process
        [Training budget
            [Small number of mcts simulations \textit{(1 SE; 0 GH)}, draw=green, top color=green!5, bottom color=green!30]
            [Not enough episodes/iterations (training) \textit{(5 SE; 1 GH)}, draw=green, top color=green!5, bottom color=green!30]]
        [RL function approximator design
            [Missing policy-gradient clipping \textit{(0 SE; 1 GH)}, draw=green, top color=green!5, bottom color=green!30]
            [Suboptimal learning start
            \textit{(0 SE; 1 GH)}, draw=green, top color=green!5, bottom color=green!30]
            [Non-random starting state \textit{(2 SE; 1 GH)}, draw=green, top color=green!5, bottom color=green!30]
            [Non-stochastic policy \textit{(0 SE; 1 GH)}, draw=green, top color=green!5, bottom color=green!30]
            [Use of inappropriate function approximator for the \\ given environment \textit{(1 SE; 0 GH)}, draw=green, top color=green!5, bottom color=green!30]]
        [Use of variable horizon in environment where fixed \\ horizon is more appropriate \textit{(0 SE; 1 GH)}, draw=green, top color=green!5, bottom color=green!30]
        [Suboptimal number of rollout steps \textit{(0 SE; 1 GH)}, draw=green, top color=green!5, bottom color=green!30]
        [Normalisation
            [Missing action normalisation \textit{(0 SE; 1 GH)}, draw=green, top color=green!5, bottom color=green!30]
            [Suboptimal / Missing normalisaton of observations \\ \textit{(1 SE; 6 GH)}, draw=green, top color=green!5, bottom color=green!30]
            [Wrong normalisation of advantage \textit{(0 SE; 1 GH)}, draw=green, top color=green!5, bottom color=green!30]]
            ]
    [Inference
        [Misconfiguration of the agent to a deterministic inference\\ \textit{(1 SE; 1 GH)}, draw=green, top color=green!5, bottom color=green!30]
        [Misconfiguration of the agent to a stochastic inference\\ \textit{(0 SE; 1 GH)}, draw=green, top color=green!5, bottom color=green!30]
        ]
] 
\end{rltaxonomy} 
\end{minipage} & 
\begin{minipage}{9cm}
\begin{rltaxonomy}
[RL Fault
    [Environment setup 
        [Termination flags 
            [Flags indicating successful termination or \\ timeout not set properly \textit{(2 SE; 3 GH)}, draw=green, top color=green!5, bottom color=green!30]
            [Missing check for dones \textit{(0 SE; 1 GH)}, draw=green, top color=green!5, bottom color=green!30]
        ]
        [Action selection
            [Suboptimal action space \textit{(1 SE; 1 GH)}, draw=green, top color=green!5, bottom color=green!30]
        ]
        [State representation
            [Subtoptimal frame skip parameter \\ \textit{(1 SE; 0 GH)}, draw=green, top color=green!5, bottom color=green!30]
            [Suboptimal scaling of features \textit{(1 SE; 0 GH)}, draw=green, top color=green!5, bottom color=green!30]
            [Suboptimal features for state \\ representation \textit{(1 SE; 0 GH)}, draw=green, top color=green!5, bottom color=green!30]
            [Suboptimal state space \textit{(7 SE; 0 GH)}, draw=green, top color=green!5, bottom color=green!30]
        ]
    ]
    [Suboptimal replay buffer 
        [Suboptimal replay buffer design \textit{(2 SE; 1 GH)}, draw=green, top color=green!5, bottom color=green!30]
        [Suboptimal replay buffer size \textit{(2 SE; 0 GH)}, draw=green, top color=green!5, bottom color=green!30]
        [Suboptimal sampling from replay buffer \\ \textit{(1 SE; 0 GH)}, draw=green, top color=green!5, bottom color=green!30]
    ]
    [Network update
        [Suboptimal network update \\ frequency \textit{(3 SE; 1 GH)}, draw=orange, top color=orange!5, bottom color=orange!30]
        [Biased target network \textit{(0 SE; 1 GH)}, draw=green, top color=green!5, bottom color=green!30]
        [Suboptimal Polyak constant value \textit{(1 SE; 0 GH)}, draw=green, top color=green!5, bottom color=green!30]
        [Suboptimally balancing value and policy \\ networks \textit{(1 SE; 0 GH)}, draw=green, top color=green!5, bottom color=green!30]
        [Wrong network update / Wrong update rule \\ \textit{(2 SE; 1 GH)}, draw=orange, top color=orange!5, bottom color=orange!30]
    ]
    [Environment exploration 
        [Suboptimal exploration rate - Epsilon \textit{(2 SE; 0 GH)}, draw=green, top color=green!5, bottom color=green!30]
        [Suboptimal exploration decay \textit{(2 SE; 0 GH)}, draw=blue, top color=blue!5, bottom color=blue!30]
        [Missing exploration \textit{(2 SE; 0 GH)}, draw=orange, top color=orange!5, bottom color=orange!30]
        [Suboptimal minimum exploration rate \textit{(1 SE; 0 GH)}, draw=green, top color=green!5, bottom color=green!30]
        [Suboptimal application of Dirichilet noise \\ \textit{(1 SE; 0 GH)}, draw=green, top color=green!5, bottom color=green!30]
        [Suboptimal noise sampling \textit{(1 SE; 0 GH)}, draw=green, top color=green!5, bottom color=green!30]
        [Missing reset/close environment \textit{(2 SE; 3 GH)}, draw=orange, top color=orange!5, bottom color=orange!30]
    ]
    [Regularisation~\cite{andrychowicz2021matters}  
        [Suboptimal entropy coefficient \textit{(1 SE; 0 GH)}, draw=green, top color=green!5, bottom color=green!30]
    ]
] 
\end{rltaxonomy}
\end{minipage}
\end{tabular}

\caption{Taxonomy of real RL faults: green indicates new fault types; orange and blue indicate fault types in common with the previous taxonomy~\cite{nikanjam2022faults}; blue indicates the ones that we renamed. SE/GH are preceded by the number of instances found in StackExchange/Github.} \label{fig:taxonomy}

\end{figure*}

\subsection{Taxonomy Construction}
We followed Islam \etal's~\cite{islam2019comprehensive} approach to build the taxonomy tree, wherein we built our tree on top of another existing RL fault tree by Nikanjam \etal's~\cite{nikanjam2022faults} (marked in orange/blue in \autoref{fig:taxonomy}).
For new labels (marked in green in \autoref{fig:taxonomy}), we followed Humbatova \etal's~\cite{humbatova2020taxonomy} approach to group them into higher-level categories.

\subsection{The Final Taxonomy}\label{sec:taxonomy}

\textbf{Reward Function.}
This category is related to faults affecting the reward function guiding the RL agent towards the learning objective.

\noindent
\ding{182} \textit{Defining the reward function} - Designing the reward function is critical to achieve good performance in RL. We found the following faults associated with it. 
\textit{Suboptimal reward function} -- Defining a good reward function is challenging, especially for complex tasks involving multiple constraints. For instance, learning a robust policy for quadrupedal
robots, requires a complex reward function encouraging linear and angular velocities, while penalising vibrations and energy consumption~\cite{takahiro2022locomotion}.
Manually setting weights for these components is not an easy task for developers~\cite{abbeel2004apprenticeship}, while they may greatly impact the learning effectiveness.  
\textit{Sparse Reward} -- This deals with cases where the reward is provided on rare occasions, for instance, at the end of each episode rather than at each timestep. In some environments, a sparse reward makes training ineffective or even impossible~\cite{riedmiller2018learning}. 
\textit{Reward function not defined for the entire range of behavior} -- This fault occurs when the reward function does not account for all of the trajectories that the agent might take. \textit{Potential of terminal step not fixed to zero} -- This fault is related to the process of reward shaping, wherein the agent is provided with supplemental rewards, to make the learning smoother. When the shaping function is based on a potential, the value of such potential should be zero at the terminal step.

\ding{183} \textit{Normalising the reward function} -- A class of faults associated with reward functions is related to normalisation. \textit{Missing reward normalisation/scaling} -- Reward scaling typically involves taking the product of the environment rewards with a scalar $(\hat{r} = r\hat{\sigma})$~\cite{duan2016benchmarking, gu2016q}. 
In certain environments clipping is an alternative to scaling. 
Existing studies~\cite{henderson2018deep} show that the choice of reward scaling/clipping has a large impact on the output of the training process. 
\textit{Unnecessary reward normalisation} -- This fault occurs when reward function normalisation is not required and its use is actually detrimental to training. %
\textit{Wrapping reward after observation normalisation} -- In this fault, observations are normalised before a wrapper is applied to the reward function, making the wrapper sub-optimal.  
\textit{Suboptimal discount factor} -- The discount factor $\gamma$ is a critical parameter used to trade off future and immediate rewards. When $\gamma$ is close to 0, the agent focuses on actions that maximise the short-term reward, whereas when $\gamma$ is close to 1, the agent privileges actions that maximise future rewards. 

\textbf{Training Process.} This category consists of faults that affect the training process of the RL agent.

\ding{184} \textit{Training budget} -- This category concerns faults related to the number of iterations used to train the RL agent. \textit{Small number of mcts simulations} -- This fault was observed in the context of the AlphaGoZero algorithm~\cite{silver2017mastering}. This algorithm uses Monte Carlo Tree Search (MCTS) to learn optimal actions. Having a low number of MCTS simulations may lead to suboptimal actions being selected~\cite{silver2017mastering, coulom2006efficient, kocsis2006bandit}.
\textit{Not enough episodes/iterations (training)} -- This fault occurs when the number of training iterations for the RL algorithm is low. This prevents the RL algorithm from learning a good policy.

\ding{185} \textit{RL function approximator design} -- A critical element in RL is the function approximator learnt during training. %
This category consists of faults that occur due to the design choices related to the selection of the function approximator. 
\textit{Missing policy-gradient clipping} -- Incorporating policy-gradient clipping improves the performance of actor-critic algorithms~\cite{andrychowicz2021matters}. 
\textit{Suboptimal learning starts} -- When the training process starts, the agent is allowed to take a series of random actions without learning. The ``learning starts'' hyperparameter is a critical parameter that controls when the agent starts learning, after the training process has begun. 
\textit{Non-random starting state} -- Starting at the same state each time the agent is reset, prevents it from exploring the surrounding states and leads to overfitting. 
\textit{Non-stochastic policy} -- Certain RL algorithms require a stochastic policy and therefore the function approximator must be stochastic in nature. Implementing a deterministic function approximator could lead to a drop in performance. 
\textit{Use of inappropriate function approximator for the given environment} -- RL problems of different sizes, in terms of state and action spaces, require different approximation techniques. Relatively smaller problems might be solved using tabular methods whereas larger problems might require linear or non-linear function approximators (such as neural networks)~\cite{sutton2018reinforcement}. 

\textit{Use of variable horizon in environment where fixed horizon is more appropriate} -- This fault occurs when reward learning is adopted during RL training. For effective reward learning (e.g., from human preferences), a fixed episode length was found to be often a better choice~\cite{christiano2017deep}. 

\textit{Suboptimal number of rollout steps} -- This fault occurs in the context of on-policy algorithms and refers to the number of rollout steps per environment used to update the policy. This hyperparameter significantly affects the algorithm's performance~\cite{andrychowicz2021matters}.

\ding{186} \textit{Normalisation} -- This category is related to normalisation in the context of training. \textit{Missing action normalisation} -- Action normalisation has been found to be helpful, especially when the actions are continuous~\cite{he2021computational}.  \textit{Suboptimal / Missing normalisaton of observations} -- Andrychowicz \etal~\cite{andrychowicz2021matters} recommend to always normalise observations. As per their experiments, doing this was critical for achieving high performance in almost all the environments. \textit{Wrong normalisation of advantage} -- RL algorithms such as PPO~\cite{schulman2017proximal} and A3C~\cite{mnih2016asynchronous}, compute the advantage function, i.e., an estimate of the value of a certain action in a given observation. Normalising this estimate with a single sample or a few samples may result in diverging computations (NaN), and to gradients with large variances. 

\textbf{Inference.} This category deals with faults that occur at inference time, i.e., after the RL algorithm has been trained.

\textit{Misconfiguration of the agent to a deterministic inference} -- During inference, forcing an agent to take deterministic actions when it was trained with a stochastic policy, might lead to a drop in performance~\cite{sutton2018reinforcement}. 
In fact, for problems where appreciable generalisation is required at inference time (e.g., when there is a substantial development-to-production shift), a better policy may be a stochastic one. %
\textit{Misconfiguration of the agent to a stochastic inference} -- Forcing an agent that was trained with a deterministic policy to become stochastic at inference time, thereby carrying out exploratory behavior, can also lead to a performance drop.

\textbf{Environment setup.} %
Faults related to the environment setup fall under this category. 

\ding{187} \textit{Termination flags} -- Flags that denote when an episode has ended might be set incorrectly. \textit{Flags indicating successful termination or timeout not set properly} -- Flags for termination and timeout should only be set in terminal states. Terminal states are critical for calculating state values, and these values are recursively used to compute the values for previous states. \textit{Missing check for dones} -- During training the algorithm needs to correctly check whether a state is terminal (i.e., \textit{done}), as this determines how the targets for the optimisation problem are computed.  %

\ding{188} \textit{Action selection} -- This fault occurs when the user defines an action space that makes learning difficult or impossible (e.g., representing actions as discrete integers vs bit vectors). %

\ding{189} \textit{State representation} -- This category deals with faults associated with the definition of environment states. 
\textit{Suboptimal frame skip parameter} -- The frame skip parameter forces an action to be repeated for a specific number of frames. This parameter was found to have a significant impact, in terms of learning efficiency, in environments requiring high-frequency policies, such as Atari games and robotic applications~\cite{braylan2015frame, srinivas2017dynamic}. \textit{Suboptimal scaling of features} -- State features must be scaled appropriately, in order for an RL algorithm to learn efficiently. \textit{Suboptimal features for state representation} -- In order to speed up learning, rather than feed in raw state inputs and expect the learning algorithm to identify useful patterns, developers could use their domain knowledge and engineer the state to include relevant, possibly higher level, features. \textit{Suboptimal state space} -- The RL paradigm assumes that the environment the agent operates in, follows the Markov property, i.e., that the current state the agent perceives, summarises all the past interactions of the agent with the environment. In other words, all the information the agent needs to make optimal actions, need to be in the state space of the agent. If some crucial information are hidden from the agent, the Markov property does not hold, and the agent cannot learn optimally~\cite{singh1996reinforcement}.

\textbf{Suboptimal replay buffer.} Off-policy RL algorithms typically use a replay buffer during training.

\textit{Suboptimal replay buffer design} -- Catastrophic forgetting~\cite{rolnick2019experience} might be caused by the under-representation of data for specific tasks in the replay buffer. This fault is common in multi-tasks RL settings~\cite{kirkpatrick2017overcoming}, i.e., when the RL agent has multiple objectives, but also in single environments that can be decomposed in sub-objectives (or levels)~\cite{fedus2020interference}.
\textit{Suboptimal replay buffer size} -- The replay buffer size is a non-trivial tunable hyperparameter. While a smaller replay buffer may lead to relevant data getting replaced too quickly, a large buffer might lead to older and irrelevant data getting sampled, reducing learning efficiency~\cite{zhang2017deeper}.  \textit{Suboptimal sampling from replay buffer} -- It is crucial that the sampling process maintains the i.i.d. (identical and independently distributed) property of the data, and that the sampled data are not temporally correlated. If this property does not hold, the learning process might be negatively affected.

\textbf{Network update.} This category refers to updating the parameters of the neural networks implementing the function approximators.

\textit{Suboptimal network update frequency} -- %
The frequency of the target network updates is too low/high, impacting the learning effectiveness~\cite{nikanjam2022faults}.
\textit{Biased target network} -- This fault occurs when the target network parameters are not independent from the online network's. The target network prevents the policy from exploring alternative solutions while the online network is being updated~\cite{mnih2015human}. This fault prevents the target network from converging to the optimal one. 
\textit{Suboptimal Polyak constant value} -- An alternative to duplicating the online network weights as target network weights, is to perform soft updates by Polyak averaging. 
The critical hyperparameter controlling such soft updates is the Polyak update coefficient~\cite{mnih2015human, lillicrap2015continuous}.  
\textit{Suboptimally balancing value and policy networks} --  This fault occurs while using the AlphaGo algorithm~\cite{silver2016mastering}. This algorithm uses MCTS to select actions by utilising value and policy networks. %
The parameter $\lambda$ balances the decisions of these two networks. A fault occurs when a poor value of the hyperparameter $\lambda$ is set~\cite{coulom2006efficient, kocsis2006bandit,silver2016mastering}.  
\textit{Wrong network update / Wrong update rule} -- %
New data cannot be optimally learned by the RL algorithm (e.g., because the learning rate of the neural networks decays too quickly)~\cite{nikanjam2022faults}.%

\textbf{Environment Exploration.} %
We found a variety of critical exploration hyperparameters in various RL algorithms. %
Exploring too little may cause the RL algorithm to be unable to discover high reward states and actions; exploring too much prevents the agent from exploiting what it has learned.

\textit{Suboptimal exploration rate - Epsilon} -- This hyperparameter refers to the suboptimal setting of the epsilon parameter, present in various RL algorithms~\cite{mnih2015human}.
\textit{Suboptimal exploration decay} -- During the start of RL training, the algorithm is expected to explore states extensively, to identify promising states and actions. However, as the algorithm progresses, the amount of exploration is typically reduced so that the agent can exploit its existing knowledge.
\textit{Missing exploration} -- This label refers to the case where the agent does not explore at all~\cite{nikanjam2022faults}.
\textit{Suboptimal minimum exploration rate} -- Once the exploration parameter has been completely decayed, it should be left to a value that is still greater than zero. This ensures that the agent continues to explore for the remaining training budget. However, too large values will interfere with learning, while a value that's too low will prevent experiencing new states and actions. \textit{Suboptimal application of Dirichilet noise} -- Dirichilet noise is used  by the AlphaGo algorithm for  exploration. The noise sampled from the Dirichilet distribution, which requires careful setting, is added to the root node's prior probabilities~\cite{silver2017mastering}.
\textit{Suboptimal noise sampling} -- This fault was found in the usage of the Deep Deterministic Policy Gradient (DDPG) algorithm. DDPG incorporates exploration during training by adding noise to actions. The choice of the distribution to sample the noise affects the exploration efficacy~\cite{lillicrap2015continuous}. \textit{Missing reset/close environment} -- This fault deals with forgetting to reset or to close the environment during training or inference~\cite{nikanjam2022faults}.

\textbf{Regularisation} --  Policy regularisation improves the performance of RL algorithms~\cite{liu2020regularization}. \\ %
\textit{Suboptimal entropy coefficient} -- The Asynchronous Actor Critic and PPO algorithms~\cite{schulman2017proximal, mnih2016asynchronous} incorporate the policy's entropy to the loss function, to improve exploration. Therefore the entropy coefficient hyperparameter becomes critical to control the exploration rate of the agent. 

\subsection{Comparing Prior Work with our Taxonomy} \label{sec:cmp}
Nikanjam \etal~\cite{nikanjam2022faults}'s final taxonomy has 11 fault categories. Our taxonomy contains five of these categories (with orange background in \autoref{fig:taxonomy}), plus one which we renamed (with blue background in \autoref{fig:taxonomy}). The remaining five categories do not match any category in our taxonomy for at least one of the following reasons: (1)~the fault could not be mapped to an RL framework, i.e., it only affects re-implementations of RL algorithms; (2)~the fault is not RL-specific, e.g., the fault is a generic DL fault; (3)~there is no evidence for the fault in our mined posts, e.g., the associated posts contain a how-to question, instead of describing actual issues and discussing possible solutions; (4)~the fault is a generic coding error; (5)~the associated post does not refer to any code implementing the RL agent. For the matching fault types, we used the same labels as Nikanjam \etal~\cite{nikanjam2022faults}, except for ``Suboptimal exploration rate'', which we renamed to ``Suboptimal exploration decay'', specifying more precisely that the fault is related to how fast/slow the exploration rate is decayed over time.

\section{Mutation Analysis}\label{mut-analysis}

\subsection{Mutation Operators}\label{sec:mut-op}

\begin{table}[htb]
	\centering
    \caption{List of proposed mutation operators in \tool.}
    \label{table:mutation-analysis:operators}
    \setlength{\tabcolsep}{1.5pt}
    \renewcommand{\arraystretch}{1}

        \begin{tabular}{lllc}
            \toprule

            \textbf{Group} & 
            \multicolumn{1}{l}{\textbf{Operator}}& \textbf{ID} & 
            \textbf{IS} \\

            \midrule
            
            \multirow{3}{*}{Reward function} & Change Discount Factor & SDF & Y \\
            & Make  Reward Sparse & SPR & N \\
            & Change Reward Scale & SRS & N \\

            \midrule
            
            \multirow{6}{*}{Training process} & Change Number of Rollout Steps & SNR & Y \\
            & Change Learning Start & SLS & Y \\
            & Reduce Episodes/Iterations & NEI & Y \\
            & Introduce Deterministic Start State & NRS & N \\
            & Remove Normalisation of Actions & MNA & N \\
            & Remove Normalisation of Observations & MNO & N \\

            \midrule
            
            Regularisation & Change Entropy Coefficient & SEC & Y \\

            \midrule
            
            \multirow{2}{*}{Network update} & Change Network Update Frequency & SNU & Y \\
            & Change Polyak Constant Value & SPV & Y \\
            
            \midrule
            
            Suboptimal Replay Buffer & Change Replay Buffer Size & SBS & N \\

            \midrule
            
            \multirow{2}{*}{Environment exploration} & Change Minimum Exploration Rate & SMR & Y \\
            & Change Exploration Rate & SER & N \\
            
            \bottomrule
        \end{tabular}
\end{table}

To define a set of mutation operators, we analysed all of the 48 unique fault types in the RL taxonomy of real faults (see \autoref{sec:taxonomy}). The extraction of mutation operators was organised into three stages. First, two of the authors independently went through the whole list of faults types and each derived potential mutation operators (MOs) from the inspected faults. Then, they have performed conflict resolution between themselves, and produced a list of proposed operators. At the next step, two other authors have separately inspected the set of candidate MOs and the faults that did not inspire any MO. Both of the authors have shown full agreement with the initial list of the operators, i.e. have not proposed any new MOs or rejected the existing ones.
At the last stage, two authors, one from each stage, have gone through the list of MOs to document their feasibility and applicability scenarios. The MO extraction process, as well as the comments on the possible implementation approaches, are available in our replication package~\cite{muPRL-repo}. In total, we  propose 15 mutation operators, with 8 of them implemented in our tool \tool. 

\autoref{table:mutation-analysis:operators} enlists the final set of proposed MOs, which are divided according to the corresponding top category of the taxonomy  (Column~1). Column~2 provides a short description of each MO, while Column~3 specifies a short abbreviated name. The last column ``IS'', which stands for ``Implementation Status'', shows whether an operator is implemented or not. The operators that are domain specific, i.e., that require a custom implementation for each case study, have not been implemented, as they are not generally applicable. For instance, the ``Make Reward Sparse'' operator requires knowledge of how the reward function is implemented in a given environment, while the ``Missing Normalisation'' operators are not applicable in environments where actions are discrete or observations are not normalised by default.

In total, the operators span six out of the eight top categories of the taxonomy. ``Training process'' is the most populated category with six of the proposed operators. Most of the operators stem from one  fault type in the taxonomy, with the name of the MO corresponding to the name of the taxonomy leaf. ``Change reward scale'' is an exception to this rule as it corresponds both to the ``Missing reward normalisation/scaling'' and ``Unnecessary reward normalisation'' fault types.

\subsection{Mutation Analysis Procedure}\label{sec:mut-procedure}

To ensure reliable and statistically sound evaluation of the quality of test sets, we adopt the definition of \textit{statistical killing}~\cite{JahangirovaICST20}: a mutant is \textit{killed} by a test set if the prediction accuracies of original and mutated model computed on such test set differ in a statistically significant way. 

However, RL presents numerous differences w.r.t. DL models that we need to account for when evaluating an RL agent.
In RL, since the agent is trained online, a test can be represented as an initial configuration of the environment where the agent operates~\cite{uesato2018rigorous,biagiola2024testing}. 
Let us consider the \textit{CartPole} subject environment, consisting of a cart moving to keep a pole vertically aligned (this is the classical inverted pendulum problem). Its initial configuration $e$ is a 4-dim vector $e = [x, \dot{x}, \theta, \dot{\theta}]$, where $x$ is the initial position of the cart, $\dot{x}$ is the initial velocity of the cart, $\theta$ is the initial angle of the pole w.r.t. the vertical axis, and $\dot{\theta}$ is the initial angular velocity of the pole. 
Trained RL agents are typically evaluated using a set of randomly generated initial environment configurations~\cite{openai-gym-leaderboard}, to test their generalisation capabilities. As there is no explicit test set to evaluate RL agents in a given environment, we refer to a test environment generator (or test generator $\textit{TG}$ for short), rather than a test set; a random $\textit{TG}$, which randomly generates initial environment configurations, is one example of $\textit{TG}$.

Let $A$ be the RL agent under test. To support statistical analysis of mutation killing~\cite{JahangirovaICST20}, we train $n$ instances of $A$ for $N$ time steps each. For each instance, an arbitrary random seed is chosen for the generation of a reproducible sequence of \textit{initial environment configurations} (environment configurations, for short), which are  used to train the agent, within the $N$ time steps budget. 
Then, for any given mutation operator $P$ and its configuration $P(j), \, j \in J_P$, we train $n$ mutant instances by reusing the same set of random seeds, and, as a result, the same sequence of initial environment configurations used to train the original agent instances. In this way, we create $n$ pairs of original and mutant instances $(o_i, m_i)$ that are trained on the same sequence (or sequence prefix, if a shorter sequence is used) of initial environments.

\subsubsection{Killability}
We first define the \textit{killed} predicate for an MO parameter configuration $P(j)$ and then we use it to define the notion of \textit{killable} (and its complement, \textit{likely-equivalent}) MO configuration $P(j)$. 

To decide whether a mutant is killed by a test generator $\textit{TG}$, we execute each pair $(o_i, m_i)$ of original and mutant agents in test mode on the sequences of environment configurations generated by $\textit{TG}$. 
Since $\textit{TG}$ might generate different sequences depending on the agent under test, with no loss of generality we assume that two different test sequences $T_{o_i} = \textit{TG}(o_i)$ and $T_{m_i} = \textit{TG}(m_i)$ are obtained when applying $\textit{TG}$ to either the original agent $o_i$ or the mutant $m_i$. When such a dependency does not hold (i.e., the dependency between $\textit{TG}$ and the agent under test), there is a single test sequence $T = T_{o_i} = T_{m_i}$. This happens for instance when using a random $\textit{TG}$ or a predefined sequence of environment configurations $T$ as test set. It is also convenient to assume that the two test sequences have the same length $L = |T_{o_i}| = |T_{m_i}|$.

We represent the result of the execution of each pair $(o_i, m_i)$ on the corresponding sequences $(T_{o_i}, T_{m_i})$ as a 4-tuple $(S_{o_i}, F_{o_i}, S_{m_i}, F_{m_i})$, where $S_{o_i}$ and $F_{o_i}$ are the number of successes and failures for the $i$-th original agent instance $o_i$, with $S_{o_i} + F_{o_i} = L$; the variables $S_{m_i}$ and $F_{m_i}$ store these measurements for the paired $i$-th mutant instance $m_i$, with $S_{m_i} + F_{m_i} = L$. Given the contingency table $[[S_{o_i}, F_{o_i}], [S_{m_i}, F_{m_i}]]$ for each pair $(o_i, m_i)$ we apply the Fisher's test~\cite{Fisher1992} to decide whether the mutant instance $m_i$ is killed or not, the \textit{killed} predicate $K$ being defined as: $K(o_i, m_i) = 1 \Leftrightarrow p_{value} < 0.05$. Note that mutating the original agent may result in a mutant that improves over the performance of the original (\textit{weaker}) agent~\cite{kim2023repairing}; in this case a certain pair can be killed because $F_{o'_i} > F_{m'_i}$, i.e., the original instance $o'_i$ fails more often than the mutated instance $m'_i$. All such pairs $(o'_i, m'_i)$ are discarded and for them the killed predicate is conventionally set to zero, i.e., $K(o'_i, m'_i) = 0$.

The \text{killed} predicate $K$ of a given mutant configuration $P(j)$ is calculated   for a given test generator $\textit{TG}$ based on the number of killed instance pairs over the total number of pairs $n - w$, where $w$ is the number of pairs where the original instance is weaker (i.e., fails more often) than its mutated instance pair:

\begin{equation*}
K(\textit{TG}, P, j)  = 
\begin{cases}
\textit{1 (true)} &\text{if}\: K_R(\textit{TG}, P, j) \geq 0.5 \\
\textit{0 (false)} &\text{otherwise}
\end{cases}%
\end{equation*}

\noindent
where $K_R(\textit{TG}, P, j) = 1/(n - w) \cdot \sum_{i=1}^{n} K(o_i, m_i)$ is the \textit{killing rate} (i.e., proportion of killed mutant instances).
A certain mutant configuration $P(j)$ is \textit{killed} if at least half of its pairs is killed, excluding those pairs where the original instance is weaker than the mutant instance. 

The notion of \textit{likely-equivalent} (and its complement, \textit{killable}) mutant that we use in our mutation procedure, is based on the one proposed in DeepCrime~\cite{deepcrime}: a mutant is \textit{likely-equivalent} if the training data cannot capture the differences between the original and mutant model. 
Hence, to decide whether the MO parameter configuration $P(j)$ is \textit{killable} we check if it is killed by the training data. Specifically, we set $\textit{TG} = \textit{TRS}_{o_i}$, i.e., we replay both $o_i$ and $m_i$ on the set of training environment configurations $\textit{TRS}_{o_i}$ and compute the killing predicate $K(\textit{TRS}_{o_i}, P, j)$. As $\textit{TRS}_{o_i}$ was used to train $o_i$ and at least a prefix of $\textit{TRS}_{o_i}$ was used to train $m_i$, we expect $\textit{TRS}_{o_i}$ to be highly discriminative between original and mutant~\cite{deepcrime}.

A mutation operator $P$ is \textit{killable} if at least one of its mutant configurations $P(j)$ is killable. 
If a mutation operator is not killable, it is deemed likely-equivalent and discarded.

\subsubsection{Triviality}
We are also interested in checking whether a certain MO generates mutants that are trivial to kill. To evaluate triviality of each mutant, we reuse the results of the replay of each original agent and mutant pair $(o_i, m_i)$ on their set of training environment configurations $\textit{TRS}_{o_i}$ from killability analysis. From $\textit{TRS}_{o_i}$, we select the subset of environment configurations where the original agent instance $o_i$ succeeds, and check in how many of these environment configurations the mutant instance $m_i$ fails. We calculate the average proportion of failing environment configurations over all the pairs, and, if it exceeds 90\%, we consider the mutant to be trivial. We exclude trivial mutants from our analysis, as they would inflate the mutation score without being discriminative.

\subsubsection{Mutation Score}
Once the likely-equivalent mutants are sorted out, for each pair $(o_i, m_i)$ we generate $L$ test environment configurations using the given test generator $\textit{TG}$. 
The mutation score of a test generator $\textit{TG}$, for a given mutation operator $P$, is the average killing rate $K_R$ across all mutant configurations $P(j)$\footnote{Considering the killing rate $K_R$ rather than the killed predicate $K$, ensures that the mutation score computation is more fine-grained.}. Given an RL mutation tool, the overall mutation score $\textit{MS}$ for a test generator $\textit{TG}$ is calculated as the average across the tool's MOs:
\begin{equation} \label{eq:mutation-score}
    \textit{MS}(\textit{TG}, \textit{OP})  = \frac {1}{|\textit{OP}|} \sum_{P \in \textit{OP}} \frac {1}{|J_P|}\sum_{j \in J_P} K_R(\textit{TG}, P, j)
\end{equation}

\section[Evaluation]{Empirical evaluation} \label{sec:evaluation}

\noindent
\textbf{RQ\textsubscript{1} [Usefulness]}: \textit{Are \tool's mutation operators useful? Do they discriminate between test environment generators of different qualities?}

In the first research question, we investigate whether the mutation testing pipeline in \tool is able to generate \textit{non-trivial, killable} and \textit{discriminative} mutants, i.e., mutants that would tell apart test environment generators of different qualities. %

\head{Metrics} To answer RQ\textsubscript{1} we measure triviality and killability for each mutation operator in each subject environment and RL algorithm. We also measure the mutation scores of two test generators (details provided in \autoref{sec:proc}), namely \textit{Weak} ($\textit{TG}_W$) and \textit{Strong} ($\textit{TG}_S$). To evaluate the discriminative power of the mutants, we measure the \textit{sensitivity} between the Weak and Strong test generators as defined in DeepCrime~\cite{deepcrime}, when $\textit{MS}(\textit{TG}_S, \textit{OP}) \geq \textit{MS}(\textit{TG}_W, \textit{OP})$:
\begin{equation} \label{eq:sensitivity}
    \textit{Sensitivity} = \frac{|\textit{MS}(\textit{TG}_S, \textit{OP}) - \textit{MS}(\textit{TG}_W, \textit{OP})|}{ \textit{MS}(\textit{TG}_S, \textit{OP})}
\end{equation}

\noindent
while we set it to zero when $\textit{MS}(\textit{TG}_S, \textit{OP}) < \textit{MS}(\textit{TG}_W, \textit{OP})$.

\noindent
\textbf{RQ\textsubscript{2} [Comparison]}: \textit{How does \tool compare with the state-of-the-art RLMutation approach?}

In this research question, we compare our tool with an existing mutation tool for RL, namely RLMutation~\cite{TambonRLMut}. 

\head{Metrics} To answer RQ\textsubscript{2}, we compare \tool and RLMutation on the same subject environments and RL algorithms used for RQ\textsubscript{1}, by measuring sensitivity of $\textit{TG}_W$ and $\textit{TG}_S$ on the mutants produced by both approaches.

\noindent
\textbf{RQ\textsubscript{3} [New Fault Types]}: \textit{What is the impact of the new fault types identified in our taxonomy and implemented in \tool?}

In this research question, we evaluate the specific contribution of the new fault types that emerge from our taxonomy, w.r.t. existing RL fault taxonomies in the literature. In particular, we consider the impact of the five new mutation operators, namely SNR, SLS, NEI, SEC, and SPV.

\head{Metrics} To answer RQ\textsubscript{3}, we compute the sensitivity of the newly introduced fault types for each pair subject environment ($\textit{env}$) -- RL algorithm ($A$). %

\subsection{Subject Environments and RL Algorithms} \label{sec:subjects}

\head{Subject Environments} We evaluated our approach using two subject environments used in previous work~\cite{TambonRLMut}, namely \textit{CartPole}~\cite{barto1983neuronlike} and \textit{LunarLander}~\cite{lunarlander} to be able to compare our approach with RLMutation, and we added two new subject environments, one concerning the driving domain, i.e., \textit{Parking}~\cite{LeurentAnEnvironment2018}, and a robotic environment, namely \textit{Humanoid}~\cite{tassa2012synthesis}, both of which are commonly used in the RL literature. Each environment has an initial configuration that is generated randomly at the beginning of each episode, according to the constraints determined by each environment.

\head{RL Algorithms} We selected four RL algorithms that are widely used in the literature. DQN~\cite{mnih2015human} is representative of value-based algorithms, while PPO~\cite{schulman2017proximal} is a widely used policy-gradient algorithm. SAC~\cite{haarnoja2018soft} and TQC~\cite{kuznetsov2020controlling} are \textit{hybrid} algorithms, i.e., blending value-based and policy-gradient techniques. DQN, SAC and TQC are off-policy algorithms, while PPO is an on-policy algorithm. The different characteristics of these four RL algorithms, allow for the application of all MOs of \tool.

\begin{table*}[ht]
	\centering
    \scriptsize
    \caption{Results for RQ\textsubscript{1}, RQ\textsubscript{2}, RQ\textsubscript{3}. Gray cells indicate that the specific mutation operator is not applicable to a certain $(\textit{env}, A)$ pair. Boldfaced values indicate that an operator is killable, while the symbol ``--'' stands for ``not available''. Underlined operators indicate new fault types w.r.t. existing taxonomies, and \underline{\textit{Avg new}} refers to the average computed only for underlined operators. The sensitivity of RLMutation is reported in the last row.}
    \label{table:evaluation:rq1-rq2-rq3}
    \setlength{\tabcolsep}{1.65pt}
    \renewcommand{\arraystretch}{1.3}

        \begin{tabular}{r@{\hskip 0.5em}cccccccccccccccccccccccccccccccccccccccc}
            \toprule

            \multicolumn{1}{l}{} & \multicolumn{10}{c}{\textbf{CartPole}} & \multicolumn{10}{c}{\textbf{LunarLander}} & \multicolumn{10}{c}{\textbf{Parking}} & \multicolumn{10}{c}{\textbf{Humanoid}} \\

            \cmidrule(r){2-11}
            \cmidrule(r){12-21}
            \cmidrule(r){22-31}
            \cmidrule(r){32-41}
            
            \multicolumn{1}{l}{} & \multicolumn{5}{c}{DQN} & \multicolumn{5}{c}{PPO} & \multicolumn{5}{c}{PPO} & \multicolumn{5}{c}{DQN} & \multicolumn{5}{c}{SAC} & \multicolumn{5}{c}{TQC} & \multicolumn{5}{c}{SAC} & \multicolumn{5}{c}{TQC} \\

            \cmidrule(r){2-6}
            \cmidrule(r){7-11}
            \cmidrule(r){12-16}
            \cmidrule(r){17-21}
            \cmidrule(r){22-26}
            \cmidrule(r){27-31}
            \cmidrule(r){32-36}
            \cmidrule(r){37-41}

            \multicolumn{1}{l}{} & \multicolumn{1}{l}{} & \multicolumn{1}{l}{} & \multicolumn{2}{c}{MS} & \multicolumn{1}{l}{} & \multicolumn{1}{l}{} &  & \multicolumn{2}{c}{MS} & \multicolumn{1}{l}{} & \multicolumn{1}{l}{} &  & \multicolumn{2}{c}{MS} & \multicolumn{1}{l}{} & \multicolumn{1}{l}{} & \multicolumn{1}{l}{} & \multicolumn{2}{c}{MS} & \multicolumn{1}{l}{} & \multicolumn{1}{l}{} &  & \multicolumn{2}{c}{MS} & \multicolumn{1}{l}{} & \multicolumn{1}{l}{} &  & \multicolumn{2}{c}{MS} & \multicolumn{1}{l}{} & \multicolumn{1}{l}{} &  & \multicolumn{2}{c}{MS} & \multicolumn{1}{l}{} & \multicolumn{1}{l}{} &  & \multicolumn{2}{c}{MS} & \multicolumn{1}{l}{} \\

            \cmidrule(r){4-5}
            \cmidrule(r){9-10}
            \cmidrule(r){14-15}
            \cmidrule(r){19-20}
            \cmidrule(r){24-25}
            \cmidrule(r){29-30}
            \cmidrule(r){34-35}
            \cmidrule(r){39-40}
            
            \multicolumn{1}{l}{} & 
            \rot{\% trivial} & 
            \rot{\% killable}
            & \rot{Weak}
            & \rot{Strong}
            & \rot{Sensitivity}
            & \rot{\% trivial}
            & \rot{\% killable}
            & \rot{Weak} 
            & \rot{Strong} 
            & \rot{Sensitivity}
            & \rot{\% trivial}
            & \rot{\% killable}
            & \rot{Weak} 
            & \rot{Strong} 
            & \rot{Sensitivity}
            & \rot{\% trivial}
            & \rot{\% killable}
            & \rot{Weak} 
            & \rot{Strong} 
            & \rot{Sensitivity}
            & \rot{\% trivial}
            & \rot{\% killable}
            & \rot{Weak} 
            & \rot{Strong} 
            & \rot{Sensitivity}
            & \rot{\% trivial}
            & \rot{\% killable}
            & \rot{Weak} 
            & \rot{Strong} 
            & \rot{Sensitivity}
            & \rot{\% trivial}
            & \rot{\% killable}
            & \rot{Weak} 
            & \rot{Strong} 
            & \rot{Sensitivity}
            & \rot{\% trivial}
            & \rot{\% killable}
            & \rot{Weak} 
            & \rot{Strong}
            & \rot{Sensitivity} \\
     
            \midrule
            
            \underline{SEC} & \multicolumn{1}{l}{\cellcolor[HTML]{D9D9D9}} & \multicolumn{1}{l}{\cellcolor[HTML]{D9D9D9}} & \multicolumn{1}{l}{\cellcolor[HTML]{D9D9D9}} & \multicolumn{1}{l}{\cellcolor[HTML]{D9D9D9}} & \multicolumn{1}{l}{\cellcolor[HTML]{D9D9D9}} & .00 & .00 & -- & -- & -- & .00 & \textbf{1.0} & .10 & .78 & .87 & \multicolumn{1}{l}{\cellcolor[HTML]{D9D9D9}} & \multicolumn{1}{l}{\cellcolor[HTML]{D9D9D9}} & \multicolumn{1}{l}{\cellcolor[HTML]{D9D9D9}} & \multicolumn{1}{l}{\cellcolor[HTML]{D9D9D9}} & \multicolumn{1}{l}{\cellcolor[HTML]{D9D9D9}} & \multicolumn{1}{l}{\cellcolor[HTML]{D9D9D9}} & \multicolumn{1}{l}{\cellcolor[HTML]{D9D9D9}} & \multicolumn{1}{l}{\cellcolor[HTML]{D9D9D9}} & \multicolumn{1}{l}{\cellcolor[HTML]{D9D9D9}} & \multicolumn{1}{l}{\cellcolor[HTML]{D9D9D9}} & \multicolumn{1}{l}{\cellcolor[HTML]{D9D9D9}} & \multicolumn{1}{l}{\cellcolor[HTML]{D9D9D9}} & \multicolumn{1}{l}{\cellcolor[HTML]{D9D9D9}} & \multicolumn{1}{l}{\cellcolor[HTML]{D9D9D9}} & \multicolumn{1}{l}{\cellcolor[HTML]{D9D9D9}} & \multicolumn{1}{l}{\cellcolor[HTML]{D9D9D9}} & \multicolumn{1}{l}{\cellcolor[HTML]{D9D9D9}} & \multicolumn{1}{l}{\cellcolor[HTML]{D9D9D9}} & \multicolumn{1}{l}{\cellcolor[HTML]{D9D9D9}} & \multicolumn{1}{l}{\cellcolor[HTML]{D9D9D9}} & \multicolumn{1}{l}{\cellcolor[HTML]{D9D9D9}} & \multicolumn{1}{l}{\cellcolor[HTML]{D9D9D9}} & \multicolumn{1}{l}{\cellcolor[HTML]{D9D9D9}} & \multicolumn{1}{l}{\cellcolor[HTML]{D9D9D9}} & \multicolumn{1}{l}{\cellcolor[HTML]{D9D9D9}} \\

            SMR & .20 & \textbf{.60} & .90 & .90 & .00 & \multicolumn{1}{l}{\cellcolor[HTML]{D9D9D9}} & \multicolumn{1}{l}{\cellcolor[HTML]{D9D9D9}} & \multicolumn{1}{l}{\cellcolor[HTML]{D9D9D9}} & \multicolumn{1}{l}{\cellcolor[HTML]{D9D9D9}} & \multicolumn{1}{l}{\cellcolor[HTML]{D9D9D9}} & \multicolumn{1}{l}{\cellcolor[HTML]{D9D9D9}} & \multicolumn{1}{l}{\cellcolor[HTML]{D9D9D9}} & \multicolumn{1}{l}{\cellcolor[HTML]{D9D9D9}} & \multicolumn{1}{l}{\cellcolor[HTML]{D9D9D9}} & \multicolumn{1}{l}{\cellcolor[HTML]{D9D9D9}} & .00 & \textbf{.80} & .14 & 1.0 & .86 & \multicolumn{1}{l}{\cellcolor[HTML]{D9D9D9}} & \multicolumn{1}{l}{\cellcolor[HTML]{D9D9D9}} & \multicolumn{1}{l}{\cellcolor[HTML]{D9D9D9}} & \multicolumn{1}{l}{\cellcolor[HTML]{D9D9D9}} & \multicolumn{1}{l}{\cellcolor[HTML]{D9D9D9}} & \multicolumn{1}{l}{\cellcolor[HTML]{D9D9D9}} & \multicolumn{1}{l}{\cellcolor[HTML]{D9D9D9}} & \multicolumn{1}{l}{\cellcolor[HTML]{D9D9D9}} & \multicolumn{1}{l}{\cellcolor[HTML]{D9D9D9}} & \multicolumn{1}{l}{\cellcolor[HTML]{D9D9D9}} & \multicolumn{1}{l}{\cellcolor[HTML]{D9D9D9}} & \multicolumn{1}{l}{\cellcolor[HTML]{D9D9D9}} & \multicolumn{1}{l}{\cellcolor[HTML]{D9D9D9}} & \multicolumn{1}{l}{\cellcolor[HTML]{D9D9D9}} & \multicolumn{1}{l}{\cellcolor[HTML]{D9D9D9}} & \multicolumn{1}{l}{\cellcolor[HTML]{D9D9D9}} & \multicolumn{1}{l}{\cellcolor[HTML]{D9D9D9}} & \multicolumn{1}{l}{\cellcolor[HTML]{D9D9D9}} & \multicolumn{1}{l}{\cellcolor[HTML]{D9D9D9}} & \multicolumn{1}{l}{\cellcolor[HTML]{D9D9D9}} \\
            
            SDF & .00 & \textbf{.80} & .40 & .44 & .09 & .00 & \textbf{.20} & .40 & .50 & .20 & .25 & \textbf{1.0} & .10 & .67 & .85 & .20 & \textbf{1.0} & .67 & .70 & .04 & .20 & \textbf{1.0} & .80 & 1.0 & .20 & .00 & \textbf{.80} & .10 & .60 & .83 & .75 & .75 & -- & -- & -- & .60 & \textbf{.80} & .10 & .20 & .50 \\
            
            \underline{SLS} & .40 & \textbf{.60} & .50 & .75 & .33 & \multicolumn{1}{l}{\cellcolor[HTML]{D9D9D9}} & \multicolumn{1}{l}{\cellcolor[HTML]{D9D9D9}} & \multicolumn{1}{l}{\cellcolor[HTML]{D9D9D9}} & \multicolumn{1}{l}{\cellcolor[HTML]{D9D9D9}} & \multicolumn{1}{l}{\cellcolor[HTML]{D9D9D9}} & \multicolumn{1}{l}{\cellcolor[HTML]{D9D9D9}} & \multicolumn{1}{l}{\cellcolor[HTML]{D9D9D9}} & \multicolumn{1}{l}{\cellcolor[HTML]{D9D9D9}} & \multicolumn{1}{l}{\cellcolor[HTML]{D9D9D9}} & \multicolumn{1}{l}{\cellcolor[HTML]{D9D9D9}} & .00 & \textbf{.80} & .17 & .60 & .72 & .00 & \textbf{.80} & .00 & 1.0 & 1.0 & .00 & \textbf{.60} & .00 & .30 & 1.0 & .00 & \textbf{1.0} & .00 & .22 & 1.0 & .00 & .00 & -- & -- & -- \\
            
            \underline{SNR} & \multicolumn{1}{l}{\cellcolor[HTML]{D9D9D9}} & \multicolumn{1}{l}{\cellcolor[HTML]{D9D9D9}} & \multicolumn{1}{l}{\cellcolor[HTML]{D9D9D9}} & \multicolumn{1}{l}{\cellcolor[HTML]{D9D9D9}} & \multicolumn{1}{l}{\cellcolor[HTML]{D9D9D9}} & .00 & .00 & -- & -- & -- & .00 & \textbf{.75} & .10 & .75 & .87 & \multicolumn{1}{l}{\cellcolor[HTML]{D9D9D9}} & \multicolumn{1}{l}{\cellcolor[HTML]{D9D9D9}} & \multicolumn{1}{l}{\cellcolor[HTML]{D9D9D9}} & \multicolumn{1}{l}{\cellcolor[HTML]{D9D9D9}} & \multicolumn{1}{l}{\cellcolor[HTML]{D9D9D9}} & \multicolumn{1}{l}{\cellcolor[HTML]{D9D9D9}} & \multicolumn{1}{l}{\cellcolor[HTML]{D9D9D9}} & \multicolumn{1}{l}{\cellcolor[HTML]{D9D9D9}} & \multicolumn{1}{l}{\cellcolor[HTML]{D9D9D9}} & \multicolumn{1}{l}{\cellcolor[HTML]{D9D9D9}} & \multicolumn{1}{l}{\cellcolor[HTML]{D9D9D9}} & \multicolumn{1}{l}{\cellcolor[HTML]{D9D9D9}} & \multicolumn{1}{l}{\cellcolor[HTML]{D9D9D9}} & \multicolumn{1}{l}{\cellcolor[HTML]{D9D9D9}} & \multicolumn{1}{l}{\cellcolor[HTML]{D9D9D9}} & \multicolumn{1}{l}{\cellcolor[HTML]{D9D9D9}} & \multicolumn{1}{l}{\cellcolor[HTML]{D9D9D9}} & \multicolumn{1}{l}{\cellcolor[HTML]{D9D9D9}} & \multicolumn{1}{l}{\cellcolor[HTML]{D9D9D9}} & \multicolumn{1}{l}{\cellcolor[HTML]{D9D9D9}} & \multicolumn{1}{l}{\cellcolor[HTML]{D9D9D9}} & \multicolumn{1}{l}{\cellcolor[HTML]{D9D9D9}} & \multicolumn{1}{l}{\cellcolor[HTML]{D9D9D9}} & \multicolumn{1}{l}{\cellcolor[HTML]{D9D9D9}} & \multicolumn{1}{l}{\cellcolor[HTML]{D9D9D9}} \\
            
            \underline{NEI} & .00 & \textbf{.80} & .60 & .67 & .10 & .00 & \textbf{.20} & .50 & .60 & .17 & .00 & \textbf{.67} & .00 & .70 & 1.0 & .20 & \textbf{.60} & .00 & .67 & 1.0 & .20 & \textbf{1.0} & .00 & .80 & 1.0 & .00 & \textbf{.80} & .00 & .30 & 1.0 & .20 & \textbf{1.0} & .00 & .20 & 1.0 & .00 & \textbf{.40} & .00 & .00 & -- \\
            
            SNU & .20 & \textbf{.80} & .90 & .80 & .00 & \multicolumn{1}{l}{\cellcolor[HTML]{D9D9D9}} & \multicolumn{1}{l}{\cellcolor[HTML]{D9D9D9}} & \multicolumn{1}{l}{\cellcolor[HTML]{D9D9D9}} & \multicolumn{1}{l}{\cellcolor[HTML]{D9D9D9}} & \multicolumn{1}{l}{\cellcolor[HTML]{D9D9D9}} & \multicolumn{1}{l}{\cellcolor[HTML]{D9D9D9}} & \multicolumn{1}{l}{\cellcolor[HTML]{D9D9D9}} & \multicolumn{1}{l}{\cellcolor[HTML]{D9D9D9}} & \multicolumn{1}{l}{\cellcolor[HTML]{D9D9D9}} & \multicolumn{1}{l}{\cellcolor[HTML]{D9D9D9}} & .00 & \textbf{1.0} & .56 & .88 & .36 & \multicolumn{1}{l}{\cellcolor[HTML]{D9D9D9}} & \multicolumn{1}{l}{\cellcolor[HTML]{D9D9D9}} & \multicolumn{1}{l}{\cellcolor[HTML]{D9D9D9}} & \multicolumn{1}{l}{\cellcolor[HTML]{D9D9D9}} & \multicolumn{1}{l}{\cellcolor[HTML]{D9D9D9}} & \multicolumn{1}{l}{\cellcolor[HTML]{D9D9D9}} & \multicolumn{1}{l}{\cellcolor[HTML]{D9D9D9}} & \multicolumn{1}{l}{\cellcolor[HTML]{D9D9D9}} & \multicolumn{1}{l}{\cellcolor[HTML]{D9D9D9}} & \multicolumn{1}{l}{\cellcolor[HTML]{D9D9D9}} & \multicolumn{1}{l}{\cellcolor[HTML]{D9D9D9}} & \multicolumn{1}{l}{\cellcolor[HTML]{D9D9D9}} & \multicolumn{1}{l}{\cellcolor[HTML]{D9D9D9}} & \multicolumn{1}{l}{\cellcolor[HTML]{D9D9D9}} & \multicolumn{1}{l}{\cellcolor[HTML]{D9D9D9}} & \multicolumn{1}{l}{\cellcolor[HTML]{D9D9D9}} & \multicolumn{1}{l}{\cellcolor[HTML]{D9D9D9}} & \multicolumn{1}{l}{\cellcolor[HTML]{D9D9D9}} & \multicolumn{1}{l}{\cellcolor[HTML]{D9D9D9}} & \multicolumn{1}{l}{\cellcolor[HTML]{D9D9D9}} \\
            
            \underline{SPV} & \multicolumn{1}{l}{\cellcolor[HTML]{D9D9D9}} & \multicolumn{1}{l}{\cellcolor[HTML]{D9D9D9}} & \multicolumn{1}{l}{\cellcolor[HTML]{D9D9D9}} & \multicolumn{1}{l}{\cellcolor[HTML]{D9D9D9}} & \multicolumn{1}{l}{\cellcolor[HTML]{D9D9D9}} & \multicolumn{1}{l}{\cellcolor[HTML]{D9D9D9}} & \multicolumn{1}{l}{\cellcolor[HTML]{D9D9D9}} & \multicolumn{1}{l}{\cellcolor[HTML]{D9D9D9}} & \multicolumn{1}{l}{\cellcolor[HTML]{D9D9D9}} & \multicolumn{1}{l}{\cellcolor[HTML]{D9D9D9}} & \multicolumn{1}{l}{\cellcolor[HTML]{D9D9D9}} & \multicolumn{1}{l}{\cellcolor[HTML]{D9D9D9}} & \multicolumn{1}{l}{\cellcolor[HTML]{D9D9D9}} & \multicolumn{1}{l}{\cellcolor[HTML]{D9D9D9}} & \multicolumn{1}{l}{\cellcolor[HTML]{D9D9D9}} & \multicolumn{1}{l}{\cellcolor[HTML]{D9D9D9}} & \multicolumn{1}{l}{\cellcolor[HTML]{D9D9D9}} & \multicolumn{1}{l}{\cellcolor[HTML]{D9D9D9}} & \multicolumn{1}{l}{\cellcolor[HTML]{D9D9D9}} & \multicolumn{1}{l}{\cellcolor[HTML]{D9D9D9}} & .00 & \textbf{.80} & .00 & 1.0 & 1.0 & .00 & \textbf{.40} & .00 & .90 & 1.0 & .00 & \textbf{.50} & .00 & .10 & 1.0 & .00 & .00 & -- & -- & -- \\
            
            \midrule
            
            \textit{Avg} & .16 & .72 & .66 & .71 & .11 & .00 & .10 & .45 & .55 & .18 & .06 & .85 & .08 & .73 & .90 & .08 & .84 & .31 & .77 & .60 & .10 & .90 & .20 & .95 & .80 & .00 & .65 & .03 & .53 & .96 & .24 & .81 & .00 & .17 & 1.0 & .15 & .30 & .05 & .10 & .50 \\
            
            \underline{Avg new} & .20 & .70 & .55 & .71 & .22 & .00 & .07 & .50 & .60 & .18 & .00 & .81 & .07 & .74 & .91 & .10 & .70 & .09 & .64 & .86 & .07 & .87 & .00 & .93 & 1.0 & .00 & .60 & .00 & .50 & 1.0 & .07 & .83 & .00 & .17 & 1.0 & .00 & .13 & .00 & .00 & -- \\

            \midrule
            
            \textit{RLMut.} & \multicolumn{1}{l}{} & \multicolumn{1}{l}{} & .90 & .90 & .00 & \multicolumn{1}{l}{} & \multicolumn{1}{l}{} & .92 & .92 & .00 & \multicolumn{1}{l}{} & \multicolumn{1}{l}{} & .83 & .83 & .04 & \multicolumn{1}{l}{} & \multicolumn{1}{l}{} & 1.0 & .82 & .00 & \multicolumn{1}{l}{} & \multicolumn{1}{l}{} & -- & -- & -- & \multicolumn{1}{l}{} & \multicolumn{1}{l}{} & -- & -- & -- & \multicolumn{1}{l}{} & \multicolumn{1}{l}{} & -- & -- & -- & \multicolumn{1}{l}{} & \multicolumn{1}{l}{} & -- & -- & -- \\

            \bottomrule
            
        \end{tabular}
\end{table*}

\subsection{Procedure} \label{sec:proc}

\head{RQ\textsubscript{1} [Usefulness]} For each subject environment we trained the original agents with the applicable RL algorithms using the hyperparameters provided by Raffin \etal~\cite{rl-zoo3}. DQN is only applicable to CartPole and LunarLander, as it only supports discrete action spaces, while SAC and TQC only support continous action spaces, hence they are only applicable on Parking and Humanoid. PPO cannot be applied on Parking as it is a goal-based environment that requires an off-policy algorithm. We discarded the PPO algorithm on Humanoid as with the default hyperparameters, the agent had a near zero success rate in repeated training instances.

We trained $n = 10$ original agents for each pair subject environment -- RL algoritm $(\textit{env}, A)$, to account for the randomness of the training process. Then, for each applicable mutation operator (MO) $P$ given the pair $(\textit{env}, A)$, we randomly sampled $j = 5$ mutant configurations. 

When designing the sampling range for each MO, we started from the corresponding search space already defined by Raffin \etal~\cite{rl-zoo3} for hyperparameter tuning, but we adjusted it to increase the chance of generating challenging mutant configurations. For categorical search spaces, concerning six out of eight MOs, we decreased by 50\% the upper and/or lower bounds of the original hyperparameter search space (for SDF we did not increase the upper bound as the original highest value 0.9999 is very close to the theoretical maximum 1.0). Three exceptions concern the SLS, NEI (not available in the original hyperparameter search space), and SNU operators, where we considered the corresponding mutation search space as relative to the training time steps budget.

After training the original agents and the corresponding mutant configurations for each pair $(\textit{env}, A)$, we replay the training environment configurations. For each mutation operator $P$, we select the configuration $P(j)$ that is killable, non-trivial, and closest to the original value. 

The next step after replay is building the \textit{Weak} $(\textit{TG}_W)$ and \textit{Strong} $(\textit{TG}_S)$ test environment generators. 
To obtain $\textit{TG}_W$, we first generated 200 test environment configurations at random and executed them on the original agent. During the execution we track the \textit{quality of control} ($\textit{QOC}$) of the trained agent, as a way to measure the \textit{confidence} of the agent during a certain episode. For instance in CartPole, the agent can fail in two ways, i.e., either if it brings the cart too far from the center (2.4$m$) or if the pole it is controlling falls beyond a certain angle (12$^\circ$). At each time step $t$ the $\textit{QOC}_t$ is the minimum between two (normalised) distances, i.e., the absolute distance between the current position of the cart and 2.4, and the absolute difference between the current angle of the pole and 12$^\circ$. 
Likewise, the $\textit{QOC}$ metric can be defined for the other subject environments. Then, for each test environment configuration we take the minimum $\textit{QOC}$ value, 
and we rank the 200 test environment configurations in descending order of $\textit{QOC}$. Finally, we select the first 50 test environment configurations, as those generated by $\textit{TG}_W$.

To obtain $\textit{TG}_S$, we resort to the approach by Uesato \etal~\cite{uesato2018rigorous} and Biagiola \etal~\cite{biagiola2024testing}, which consists of training a failure predictor on the training environment configurations, to learn failure patterns of the trained agent in order to generate challenging test environment configurations. Since our objective is to kill mutants, we train one neural network as failure predictor for each selected mutant configuration. 
Then, we use the trained neural network predictor in each mutant configuration to select 100 promising test environment configurations, where each selected test environment configuration is chosen to maximise the probability of the failure predictor out of 500 candidates generated at random.

\head{RQ\textsubscript{2} [Comparison]} To compare \tool with RLMutation, we considered all the mutants produced by RLMutation which are publicly available. Then, for each killable mutant of RLMutation, we executed the test environment configurations  generated by $\textit{TG}_W$ and $\textit{TG}_S$. For each original and mutant pair, we used RLMutation to compute the killed predicate; we then computed the mutation score for $\textit{TG}_W$ and $\textit{TG}_S$, as the ratio between the number of killed mutants and the total number of killable mutants.

\subsection{Results}\label{sec:mut-tool-eval-res}

\head{RQ\textsubscript{1} [Usefulness]} \autoref{table:evaluation:rq1-rq2-rq3} shows the evaluation of \tool for all  subject environments and RL algorithms. Rows represent the MOs, while  columns show the results of our mutation pipeline for each MO. For each pair $(\textit{env}, A)$, we report the percentage of trivial mutant configurations for each MO (\textit{\% trivial}), the percentage of killable configurations (\textit{\% killable}), the mutation scores (\textit{MS}) for the $\textit{TG}_W$ (\textit{Weak}) and $\textit{TG}_S$ (\textit{Strong}) test generators, and the individual sensitivity (\textit{Sensitivity}). For instance, in $(\textit{CartPole}, \textit{DQN})$, the SLS mutation operator (4th row), has 40\% of mutant configurations that are trivial, 60\% of configurations that are killable, while the mutation score for $\textit{TG}_W = 0.50$ and $\textit{TG}_S = 0.75$, hence  sensitivity is 0.33. We compute the mutation score for a given operator only if the operator is killable, and non-trivial. For instance, in $(\textit{CartPole}, \textit{PPO})$, the SEC operator is non-killable (\% killable $= 0.00$), while in $(\textit{Humanoid}, \textit{SAC})$, the SDF operator is killable (\% killable $= 0.75$), but all the killable configurations are trivial (\% trivial $= 0.75$), hence we do not compute the mutation score for them.

We observe that for CartPole the sensitivity is quite low, i.e., 0.11 for DQN and 0.18 for PPO. Indeed, the DQN agent in CartPole is very weak, such that even $\textit{TG}_W$ is effective at  killing mutant configurations (its mutation score is 0.66 on average, while the mutation score of $\textit{TG}_S$ is 0.71). On the other hand, the PPO agent on CartPole is very hard to kill for training environment configurations (on average the percentage of killable mutant configurations is 0.10), and, for killable configurations, $\textit{TG}_S$ has only a slight edge w.r.t. $\textit{TG}_W$. However, for the remaining subject environments, which are more complex than CartPole (i.e., these environments are harder to learn for an RL agent), the average sensitivity ranges from a minimum of 0.50 in $(\textit{Humanoid}, \textit{TQC})$ to a maximum of 1.00 in $(\textit{Humanoid}, \textit{SAC})$.

\begin{tcolorbox}[boxrule=0pt,frame hidden,sharp corners,enhanced,borderline north={1pt}{0pt}{black},borderline south={1pt}{0pt}{black},boxsep=2pt,left=2pt,right=2pt,top=2.5pt,bottom=2pt]
	\textbf{RQ\textsubscript{1} [Usefulness]}: Overall, the mutation operators of \tool are effective at discriminating strong from weak test generators, especially in complex environments where the minimum sensitivity is 0.50 and the maximum is 1.0.
\end{tcolorbox}

\head{RQ\textsubscript{2} [Comparison]} The last row of \autoref{table:evaluation:rq1-rq2-rq3} shows the average mutation score and sensitivity of RLMutation on the common pairs of subject environments and RL algorithms. We observe that, in all cases, the mutants created by \tool are more sensitive than the mutants of RLMutation, whose maximum sensitivity is 0.04, for $(\textit{LunarLander}, \textit{PPO})$.

\begin{tcolorbox}[boxrule=0pt,frame hidden,sharp corners,enhanced,borderline north={1pt}{0pt}{black},borderline south={1pt}{0pt}{black},boxsep=2pt,left=2pt,right=2pt,top=2.5pt,bottom=2pt]
	\textbf{RQ\textsubscript{2} [Comparison]}: In all subject environments and RL algorithms, \tool create mutants that are  more sensitive than RLMutation's. 
\end{tcolorbox}

\head{RQ\textsubscript{3} [New Fault Types]} In \autoref{table:evaluation:rq1-rq2-rq3} we underline the mutation operators coming from the new fault types that are not present in existing RL fault taxonomies. The \textit{Avg new} row shows the average metric values  considering only the underlined mutation operators. For instance, for the pair $(\textit{CartPole}, \textit{DQN})$, the average sensitivity across all mutation operators is 0.11, while the average sensitivity only considering the new fault types is 0.22. 
Overall, the mutation operators associated to new fault types have higher sensitivity than the total average in 5 cases; the same sensitivity in 2 cases (in one case sensitivity is not computable for the new operators).

\begin{tcolorbox}[boxrule=0pt,frame hidden,sharp corners,enhanced,borderline north={1pt}{0pt}{black},borderline south={1pt}{0pt}{black},boxsep=2pt,left=2pt,right=2pt,top=2.5pt,bottom=2pt]
	\textbf{RQ\textsubscript{3} [New Fault Types]}: Mutation operators associated to new fault types contribute substantially to increase the sensitivity of \tool.
\end{tcolorbox}

\section[Threats]{Threats to Validity}
\label{sec:THREATSTOVALIDITY}

\head{Internal Validity} 
    An internal threat to the study's validity might come from the labeling of the artifacts. We addressed this threat by having at least two labelers independently label each post. We also fixed the disagreements within labelers and used Krippendorff's Alpha to quantify the disagreements.
    Furthermore, we initially conducted pilot studies 
    to refine the labeling process. 
    An additional internal validity threat concerns the subjective bias while constructing the taxonomy tree from the generated labels.
    This was alleviated by all the authors providing feedback on the final tree.

\head{External Validity}
    An external validity threat is related to the generalizability of the bugs found on Stack Exchange and GitHub. While the sources of the bugs might be limited, we cross referenced top conferences and highly cited works, to ensure that such issues have been studied in the literature.
    The selection of conferences to identify a popular RL framework also poses an external validity threat. While we considered the top tier ML conference, i.e., ICML, considering other top ML conferences, could have given us a better picture of popular RL frameworks in the ML community. 
    Generalization might also be affeced by our choice of mining Github only considering StableBaselines3. Although StableBaselines3 is the RL framework used by the majority of the papers in the selection of conferences we considered, by not investigating other RL frameworks we might lose out on a variety of important faults.
    Lastly, an additional external validity threat is related to the limited number of subject environments we considered in the evaluation. We selected \textit{CartPole} and \textit{LunarLander} to enable comparisons with previous work. Additionally, we considered \textit{Parking} and \textit{Humanoid}, which are also heavily used in DRL research. Overall, this selection of environments, supporting both discrete and continuous action spaces, allowed us to apply four foundational DRL algorithms, namely DQN, PPO, SAC and TQC.

\head{Conclusion Validity}
    Conclusion validity threats are related to how random variations in the experiments are handled, and the inappropriate use of statistical tests. Since RL algorithms are notoriously sensitive to the random seed~\cite{henderson2018deep}, we train original and mutated agents multiple times (i.e., $n = 10$), and we used rigorous statistical tests (i.e., the Fisher's test) to decide whether the mutated agent is killed. Recently Agarwal \etal~\cite{rishabh2021edge} proposed bootstrap sampling to overcome the uncertainty given by a few-run RL training regime. We acknowledge that our experimental setting may benefit from bootstrap sampling, and by using the \textsc{rliable} library~\cite{rishabh2021edge}, we re-computed the killability predicate on all 1.7$k$ mutant instances using bootstrap sampling. In particular, we computed the probability of improvement, and estimated the confidence intervals using 2$k$ samples. We found that the killed predicate computed using bootstrap sampling agrees with the killed predicate based on Fisher’s test 88\% of the time. Moreover, when in disagreement, 65\% of the times the mutant instance is killed by the Fisher’s test, indicating a higher statistical power than bootstrap sampling. Hence, these results suggest that bootstrap sampling would bring minimal benefit to our experimental setting, although it can be used as an alternative to the Fisher's test for the killed predicate.

\section[Conclusion]{Conclusions and Future Work}
We present a taxonomy  of real RL faults. Using this taxonomy, we extracted mutation operators and implemented them in our tool \tool. We evaluated its effectiveness in discriminating strong and weak test generators on a diverse set of environments using popular RL algorithms. Our tool also achieves higher sensitivity compared to the prior work, RLMutation, with a significant contribution from the  operators that are derived from new taxonomy branches. 
\label{sec:conclusion}

\section[Data Availability]{Data Availability}
\label{sec:data-availability}

Our taxonomy labeling results are available on our replication package~\cite{benchmark-and-dataset}. We also share the  code of \tool~\cite{muPRL-repo}.

\section[Acknowledgment]{Acknowledgments}
\label{sec:Acknowledgment}
We are grateful for the help provided by Breno Dantas Cruz. We also acknowledge the ICSE '25 reviewers for their valuable feedback. 
This work relied on the grants provided by the National Science Foundation: CCF-15-18897, CNS-15-13263, CNS-21-20448, CCF-19-34884, CCF-22-23812, and NRT-21-52117. This work used Explore ACCESS at Texas High Performance Research Computing through allocation CIS240181 from the Advanced Cyberinfrastructure Coordination Ecosystem: Services \& Support (ACCESS) program, which is supported by National Science Foundation grants \#2138259, \#2138286, \#2138307, \#2137603, and \#2138296~\cite{boerner2023access}.
Matteo Biagiola was partially supported by the H2020 project PRECRIME, funded under the ERC Advanced Grant 2017 Program (ERC Grant Agreement n.~787703).

\balance
\bibliographystyle{IEEEtran}
\bibliography{paper}

\end{document}